\documentclass[journal,12pt,onecolumn]{IEEEtran} 

\usepackage{cite}
\usepackage[pdftex]{graphicx}
\usepackage{amsmath}
\usepackage{amssymb}
\usepackage{amsthm}
\usepackage{xfrac}
\usepackage{epstopdf}
\usepackage{textcomp}
\usepackage{multicol}
\usepackage{ulem}
\usepackage{color}
\usepackage{tabularx}
\usepackage{algorithm}
\usepackage{algorithmicx}
\usepackage[noend]{algpseudocode}
\usepackage{charter}
\usepackage[switch]{lineno}
\usepackage{mathtools}
\usepackage{tikz}
\usetikzlibrary{calc}

\DeclareMathOperator*{\ER}{ER}

\newtheorem{definition}{Definition}
\newtheorem{theorem}{Theorem}

\newtheorem{conjecture}{Conjecture}
\newtheorem{example}{Example}
\usepackage{url}

\begin{document}

\title{Statistical Complexity of Heterogeneous Geometric Networks}

\author{
    \IEEEauthorblockN{Keith~Malcolm~Smith\IEEEauthorrefmark{1}, Jason P. Smith\IEEEauthorrefmark{2}}\\
    \IEEEauthorblockA{\IEEEauthorrefmark{1}Department of Computer and Information Sciences, University of Strathclyde, Glasgow, UK
    \\\ keith.m.smith@strath.ac.uk}\\
    \IEEEauthorblockA{\IEEEauthorrefmark{2}Department
of Mathematics, Nottingham Trent University, Nottingham,
UK}
}

\maketitle
\begin{abstract}
Degree heterogeneity and latent geometry, also referred to as popularity and similarity, are key explanatory components underlying the structure of real-world networks. The relationship between these components and the statistical complexity of networks is not well understood. We introduce a parsimonious normalised measure of statistical complexity for networks. The measure is trivially 0 in regular graphs and we prove that this measure tends to 0 in Erd\"os-R\'enyi random graphs in the thermodynamic limit. We go on to demonstrate that greater complexity arises from the combination of heterogeneous and geometric components to the network structure than either on their own. Further, the levels of complexity achieved are similar to those found in many real-world networks. However, we also find that real-world networks establish connections in a way which increases complexity and which our null models fail to explain. We study this using ten link growth mechanisms and find that only one mechanism successfully and consistently replicates this phenomenon-- probabilities proportional to the exponential of the number of common neighbours between two nodes. Common neighbours is a mechanism which implicitly accounts for degree heterogeneity and latent geometry. This explains how a simple mechanism facilitates the growth of statistical complexity in real-world networks.

\end{abstract}

\section*{Author Summary}
A statistically complex system is one which is neither regular nor random, but contains diversity in components and structure. This departs from algorithmic complexity which describes how difficult it is to explain information, but which is maximal for uniformly random information. We provide a definition of statistical complexity for networks and propose a normalised measure which satisfies that definition. We go on to explore the relationship between statistical complexity and the two major components thought to underlie network structure-- the popularity of nodes in making connections (degree heterogeneity) and the geometric similarity of nodes. We find that the statistical complexity of real-world networks agrees with a model which combines both components. We then notice a positive relationship between the density of links in real-world networks and their statistical complexity, which is not present in our modelling. We find that we can replicate this relationship, however, by growing the network using link probability which is based on a pair of nodes' number of common neighbours. We conclude that statistical complexity is a natural by-product of uncomplicated network mechanisms, returning to the old adage that complexity arises from simplicity.

\section{Introduction}

\IEEEPARstart{C}{omplexity} is a word used often in its common meaning within various scientific disciplines to describe the size and multiplicity of facets and scales within a given real-world system. In such cases, it is often used without reference to a specific measurement, or measurements are focused on counting the numbers of such facets and scales which are apparent in that system. In computer science, complexity has two specific definitions. Firstly, computational complexity describes the shortest amount of processing time (as a function of input size) it takes to derive the desired output from an algorithm \cite{Hartmanis1971}. Secondly, Kolmogorov complexity is a measure of the complexity of information based on the size of the smallest piece of code required to derive that information as an output \cite{Kolmogorov1998}. There is no systematic way of finding such a shortest piece of code and proving that it is the shortest piece of code to compute a generic piece of information \cite{Chaitin1995}, however Kolmogorov complexity is related to measures of entropy, based on the predictability of information. For a given size of information, $n$, it is understood that randomly generated pieces of information would require the largest amount of code to be deterministically reproduced, while completely regular information, e.g. aaa$\dots$a written $n$ times, would take the least amount of code to reproduce-- ``Write `a' $n$ times". In this way there is a significant interest in framing complexity in terms of information entropy, since entropy similarly dictates a scale between regular and random structures. 

Yet, while randomly generated information may be difficult to deterministically reproduce, it is not structurally complex in a statistical sense. Indeed, the statistical properties of randomly generated information are defined a-priori and are evidently simple. This led the field of dynamical systems to lay out a different conceptual framework of complexity. In this view, a measure of complexity should go to zero for regular and random structures in the thermodynamic limit (as number of components goes to infinity), while being higher for systems presenting non-trivial and diverse correlations \cite{Huberman1986, Feldman1998}. One particularly important point of developing measures of statistical complexity is that using a scale between regular and random with complexity somewhere in the middle, a common approach from an information theoretic angle, does not allow for a useful measure of complexity itself  \cite{Huberman1986, Feldman1998}. Instead, we need a scale between the simple (regularity and randomness both having uniform generational principles) and the complex, allowing us to directly measure the extent of complexity in any given system.

When it comes to studying complexity in networks, we are concerned with the complexity of interactions-- essentially, how diverse the connectivity patterns in the network are. While others borrow from the algorithmic view of complexity \cite{Morzy2017, Zenil2018}, here we are concerned with the statistical complexity of networks. A notable early work on statistical complexity of networks introduced a measure called the network diversity score and provided a comprehensive overview of other complexity and entropy measures of networks and their limitations\cite{Streib2012}. Another work considered statistical complexity in networks from an information theoretic angle, multiplying Jensen Shannon divergence of a network with network entropy \cite{Wiedermann2017}. Neither of these works, however, provides a treatment of statistical complexity as previously described, and the measurements have limitations partly owing to lack of normalisation to network size and/or density. In this study, we establish a normalisation of Hierarchical Complexity (NHC) as a network statistical complexity measure. In contrast to the network diversity score, NHC is a parsimonious calculation of a single feature of a network, rather than the product of four different measures \cite{Smith2017a}. In contrast to the measure in \cite{Wiedermann2017} it does not require a reference graph, displays strong independence to network density and vanishes in the thermodynamic limit for Erd\"os-R\'enyi random graphs.

The hierarchy referred to here is the degree hierarchy of the network. A hierarchically complex system is one for which diversity of connectivity patterns are found across hierarchical levels (either individual degrees or ranges of degrees called tiers). Its introduction was motivated by the need to measure the complicated hierarchical networks of brain function and structure where it was expected that diverse functionality would be reflected in diverse connectivity patterns. HC has so far seen limited application in fairly small ($n<100$) macro-scale human brain networks \cite{Smith2017a, Smith2019a, Blesa2021, Valdes2021, Smith2022} and a corpus of real-world networks of varied origin ($n<5000$) \cite{Smith2019b}.

A major issue with the generalisability of the HC measure is that it is not normalised by number of nodes or number of edges. Two similarly derived networks with different numbers of nodes or edges can be expected to have different values of HC. This paper addresses this issue by introducing a normalised HC (NHC) measure. We show mathematically that this measure is bounded above by $2$ and satisfies the statistical complexity definition of being asymptotically zero for Erd\"os-R\'enyi random graphs (an appropriate equivalent to randomness in dynamical systems). We then go on to explore results of this normalisation on different types of random graphs with the two most evident structural properties relevant to real-world networks, hierarchy and geometry. Hierarchy here relates to the distribution of node fitness \cite{Caldarelli2002} or popularity \cite{Papadopoulos2012}. Geometry relates to the latent space of similarities between nodes \cite{Hoff2002, SmithA2019}. Combining hierarchy and geometry successfully captures many of the properties of real-world networks \cite{Smith2021}, but whether these properties are enough to explain the statistical complexity of networks is not known. After this we go on to explore NHC in real-world networks. Here, an unexpected relationship between NHC and density is noted. Finally, we explore explanations for this relationship by applying different kinds of link growth mechanisms based on degrees and overlap of node neighbourhoods. Combined, the results reveal non-trivial hierarchical complexity in real-world networks and we open the way for more reliable and robust applications of hierarchical complexity across network domains.

\section{Theory}
\begin{figure*}[!t]
\centering
\includegraphics[trim = 0 150 0 150, width= 6in]{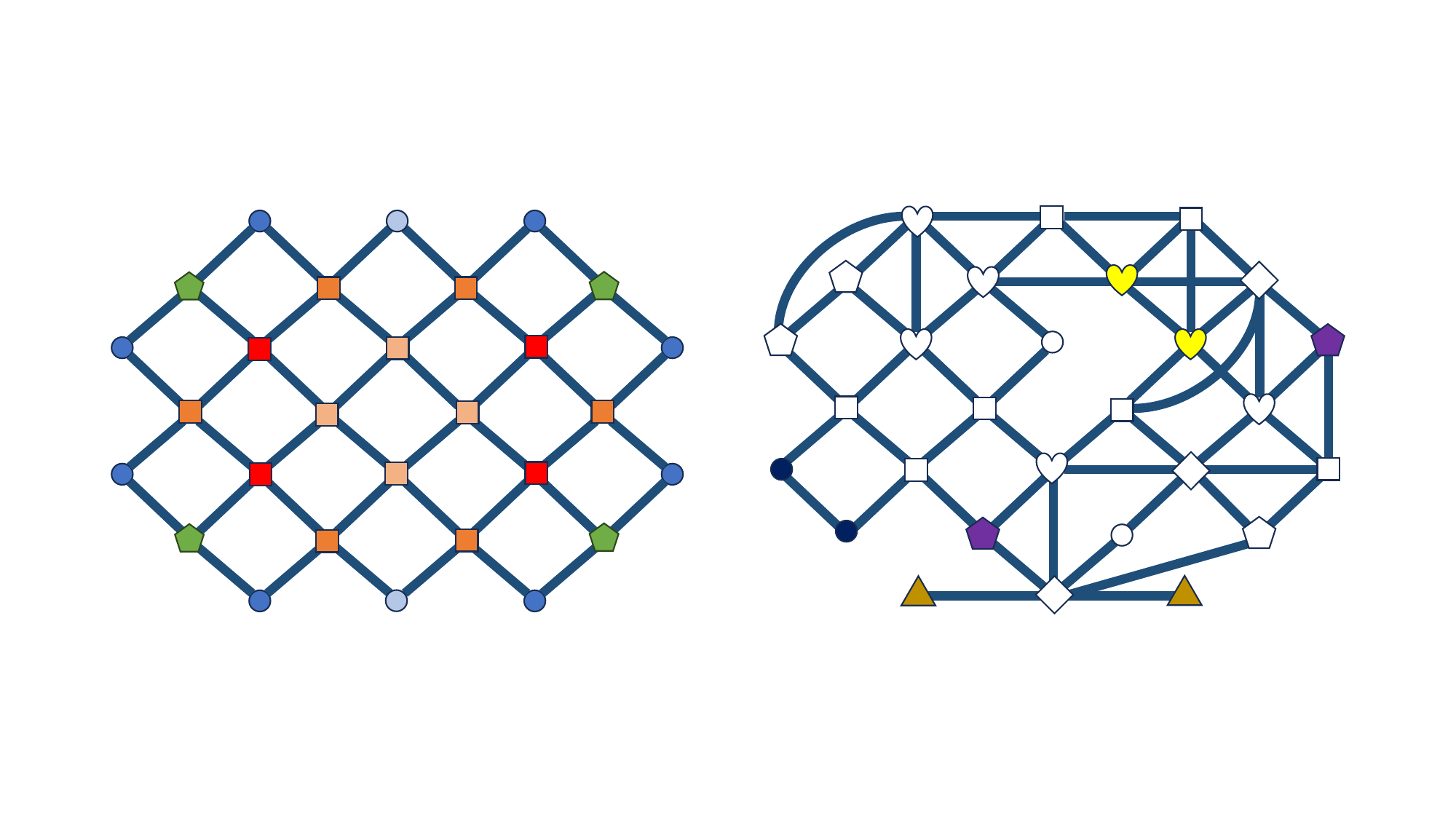}
\caption{On the left we see a geometric graph with a regular structure. Node shapes indicate distinct degrees while colours indicate distinct, repeating neighbourhood degree sequences. On the right, nodes are randomly assigned different numbers of connections. These connections are made to the closest nodes, maintaining a geometric nature, but now we see the diversity of structure this opens up. Again, different shapes indicate distinct degrees, but now there are many unique neighbourhood degree sequences which remain colourless. This diversity reflects a higher hierarchical complexity.}
\label{fig:illustration}
\end{figure*}

Key themes within this work are encapsulated within the image in Fig. \ref{fig:illustration} which illustrates hierarchical complexity arising from a combination of geometric and hierarchical structure. Henceforth, unless specified otherwise, let $G$ be a graph with $n$ nodes, $m$ edges and density $d = 2m/n(n-1)$.

\subsection{Hierarchical complexity}
To compute hierarchical complexity, we first define the \emph{Neighbourhood Degree Sequence} (NDS) of a node $i$ of degree $k$ as

\begin{equation}
    s_{i} = \{s_{i1},s_{i2},\dots,s_{ik}\} 
\end{equation}
where the $s_{ij}$'s are the degree of the nodes to which $i$ is connected such that $s_{i1}\leq s_{i2},\dots,\leq s_{ik}$. Then, for all $\ell$ nodes of degree $k$, we can stack their NDSs into an $\ell \times k$ matrix:

\begin{equation}
   \mathbf{S}_{k}(G)=
    \begin{bmatrix}
    s_{11} & s_{12} & \dots  & s_{1k} \\
    s_{21} & s_{22} & \dots  & s_{2k} \\
    \vdots & \vdots & \ddots & \vdots \\
    s_{l1} & s_{l2} & \dots  & s_{lk}
\end{bmatrix}
\end{equation}

The original definition of hierarchical complexity for degree $k$ takes the variance over the columns of this matrix and then averages across columns:
\begin{equation}\label{HC_original}
    R_{k}(G) = \frac{\sum_{j=1}^{k}\sigma_{j}^{2}}{k}
\end{equation}
where $\sigma_{j}^{2}$ is the variance of the $j$th column of $\mathbf{S}_{k}(G)$. The global measure is then the average of this over degrees:
\begin{equation}\label{hiercompeq}
    R(G) = \frac{1}{|\mathcal{D}_{2}|}\sum_{k\in\mathcal{D}_{2}} R_{k}(G).
\end{equation} 
where $\mathcal{D}_{2}$ is the set of degrees in the graph taken by at least two nodes (since variance is only meaningful over at least two elements).

\subsection{Normalised Hierarchical Complexity}

It is clear that the above definition of hierarchical complexity depends on network size. Networks of larger size have greater potential for larger degrees, which will influence the variances within equation \eqref{HC_original}. Indeed, the maximum variance of numbers in $[1,n-1]$ can occur with the sample $\{1,n-1\}$ (or any equal number of 1s and $n-1$s) which has variance $((n-2)/2)^2$. Furthermore, computations have demonstrated that hierarchical complexity also correlates with number of edges. 

Lack of normalisation to number of nodes and/or density is not unusual in network science. Indeed, the latter is essentially common place-- consider two of the most widely considered network metrics, the global clustering coefficient and global efficiency which are both maximised in complete graphs.

It is then of note that we can propose the following as a normalised measure of hierarchical complexity to both $n$ and $d$.
\begin{definition}
Let $G$ be a network, we define the \emph{$k$-th normalised hierarchical complexity} as:
\begin{equation}\label{eq:HC}\displaystyle
    \hat{R}_{k}(G) =\begin{dcases}\frac{\sum_{j=1}^{k}\sigma_{j}}{(1-d)m},&\mbox{ if $m\not=0$ and $d\not=1$}\\0,&\mbox{ if $m=0$ or $d=1$}\\\end{dcases}
\end{equation}
where $d$ is the density of $G$, $m$ is the number of edges in~$G$, and $\sigma_{j}$ is the standard deviation of the $j$th column of $\mathbf{S}_{k}(G)$, that is, the matrix where each row is the ordered degree sequence of a node of degree k. Thus we propose the normalised global measure as
\begin{equation}
\hat{R}(G) = \frac{1}{|\mathcal{D}_{2}|}\sum_{k\in\mathcal{D}_{2}}\hat{R}_{k}(G).
\end{equation}
\end{definition}

To justify this normalisation, we note that the normalised hierarchical complexity is bounded.
\begin{theorem}\label{thm:bound}
    The normalised hierarchical complexity is bounded above by $2$, that is, $\hat{R}(G)\le2$ for every graph~$G$.
    \begin{proof}
        Consider $\hat{R}_k$. Let $d_j$ be the difference of the max and min elements of the $j$-th column of $\mathbf{S}_{k}(G)$, so $\sigma_j\le \frac{d_j}{2}$. Let $\alpha = \sum_{j=1}^k d_j$.
    
        Now, let $x_j$ be the maximum degree of the $j$'th column of $\mathbf{S}_{k}(G)$. We claim that $2m\ge\sum_{j=1}^k x_j$. If $x_1,\ldots,x_k$ are degrees of distinct vertices, this follows immediately, but it could occur that $x_i$ and $x_j$ correspond to the same vertex. Note that $x_1\ge x_2\ge\dots\ge x_k$, since the rows are ordered in decreasing order. Also note that if $x_i=x_j$, with $i<j$, and $x_j$ is in row $r$, then the $i$-th entry in row $r$ must also equal $x_i$, since it cannot be smaller than something to it's right $x_j$, and cannot be bigger than the max of it's column $x_i$. And since the degrees in each row correspond to distinct vertices (as they are given by the neighourhood), if $x_i=x_j$ then there are at least two vertices of degree $x_i$. And generalising this, if $x_i$ appears as the degree of the same vertex $\ell$ times, then we can find $\ell$ distinct vertices of the same degree, so $2m\ge\sum_{j=1}^k x_j$. Also note that $x_j\ge d_j$, for all $j$, therefore
        $$2m\ge\sum_{j=1}^k x_j\ge \sum_{j=1}^k d_j=\alpha.$$
        
        Considering the minimal elements of each column and applying a similar argument to above we can also deduce that $2m'\ge\alpha$, where $m'=\frac{n(n-1)}{2}-m$ is the number of non-edges. Therefore, $\min(m,m')\ge\frac{\alpha}{2}$.
        Moreover, $\max(m,m')\ge\frac{n(n-1)}{4}$, since either at least half the edges are there or half are not. 

        So we can bound $(1-d)m$ below by:
        \begin{align*}
        (1-d)m&=\frac{2mm'}{n(n-1)}=\frac{2\max(m,m')\min(m,m')}{n(n-1)}\\&\ge\frac{(\frac{n(n-1)}{4})\alpha)}{n(n-1)}=\frac{\alpha}{4}
        \end{align*}
        Therefore, 
        $$\hat{R}_k(G)=\frac{\sum_{j=1}^{k}\sigma_{j}}{(1-d)m}\le\frac{\frac{\alpha}{2}}{(1-d)m}\le\frac{\frac{\alpha}{2}}{\frac{\alpha}{4}}=2$$
        So $\hat{R}_k(G)$ is bounded above by $2$ for all $k$, and $\hat{R}(G)$ is the average of values bounded above by $2$, hence $\hat{R}(G)\le2$.
        \end{proof}
\end{theorem}

There are several things to note about the formula \eqref{eq:HC}. Firstly, instead of variance across neighbourhood degree sequences, we here opt for standard deviation. The distribution of the standard deviation over multiple samples will in general be more symmetric than that of the variance which will be right-skewed and standard deviation is generally a more appropriate measure when normalising.

Secondly, the division by $(1-d)m$ is the term which acts to normalise the measure. This term was borrowed from the normalisation of degree variance \cite{Smith2020}. There, it was shown to normalise degree variance and bound it below 1 for all graphs.

Thirdly, instead of taking the mean over the standard deviations this normalisation takes the sum. In actuality we can consider this as a multiplication of the mean by the degree of the neighbourhood degree sequences $k$ (cancelling out the $k$ on the denominator of the average). This effectively takes account of the sampling error of taking the mean over the $\sigma_{j}$'s. The sampling error of the mean over $k$ samples is:
\begin{equation}\label{eq:samplingerror}
  \frac{\sigma}{\sqrt{k}},
\end{equation}
where $\sigma$ here is the standard deviation of the $k$ element-wise standard deviations, so that the accuracy of the mean depends on the degree $k$.

Note that in the normalisation presented we do not multiply by $\sqrt{k}$ to standardise these measurements, but by $k$ itself. This is because it is also linked with the division by $m$.

We do not believe the bound of $2$ given in Theorem~\ref{thm:bound} is tight. Through considered construction of a disconnected graph which exploits variances of 1 degree nodes, we find the largest value for $\hat{R}$ tends to $\frac{1}{3}$, which occurs with the following graph family:
\begin{example}
    Consider the graph:
    \begin{center}\begin{tikzpicture}[baseline=(current bounding box.center),line join = round, line cap = round]
    \def\c{.5}\def\s{.5}
    \node[circle, fill=black,scale=\s] (0) at (0,0*\c){};
    \node[circle, fill=black,scale=\s] (1) at (1,0*\c){};
    \node[circle, fill=black,scale=\s] (2) at (0,1*\c){};
    \node[circle, fill=black,scale=\s] (3) at (1,1*\c){};
    \node[circle, fill=black,scale=\s] (4) at (0,3*\c){};
    \node[circle, fill=black,scale=\s] (5) at (1,3*\c){};
    \node[circle, fill=black,scale=\s] (6) at (3,1.5*\c){};
    \node[circle, fill=black,scale=\s] (7) at (2,0*\c){};
    \node[circle, fill=black,scale=\s] (8) at (2,1*\c){};
    \node[circle, fill=black,scale=\s] (9) at (2,3*\c){};
    \node[circle, fill=black,scale=\s] (10) at (4,0*\c){};
    \node[circle, fill=black,scale=\s] (11) at (4,1*\c){};
    \node[circle, fill=black,scale=\s] (12) at (4,3*\c){};
    \draw[thick] (0) -- (1);\draw[thick] (2) -- (3);\draw[thick] (4) -- (5);
    \draw[thick] (7) -- (6) -- (10);
    \draw[thick] (8) -- (6) -- (11);
    \draw[thick] (9) -- (6) -- (12);
    \node[label={[rotate=-90]:$\dots$}] (dots1) at (0.4,1.8*\c){};
    \node[label={[rotate=-90]:$\dots$}] (dots2) at (1.9,1.8*\c){};
    \node[label={[rotate=-90]:$\dots$}] (dots3) at (3.9,1.8*\c){};
\end{tikzpicture}\end{center}
where each column of nodes consists of $\frac{n-1}{4}$ nodes.
 
Note that $R_1=\frac{n-3}{4}$, no other degree contributes to $R$, $m=\frac{3(n-1)}{4}$ and $d=\frac{3}{2n}$.
So $$\hat{R}=\hat{R}_1=\frac{\frac{n-3}{4}}{(1-\frac{3}{2n})\frac{3(n-1)}{4}}\rightarrow \frac{1}{3}, \,\,\,\,\,\,\,\text{ as }n\rightarrow\infty$$
\end{example}

We believe that the above family of graphs gives the largest value for $\hat{R}$, but we leave this as a conjecture:
\begin{conjecture}
    For any graph $G$ we have $\hat{R}(G)<\frac{1}{3}$.
\end{conjecture}

While we in no way claim the above family of graphs is statistically or otherwise complex, the intended application of the measure is for connected graphs with many different degrees making a contribution. We can ensure this edge case goes to zero by using the corrective term for multiply-ordered degrees as described in \cite{Smith2019b}, but for most intended purposes this is not necessary.

\subsection{Expected Complexity Values}
Next we show that $\hat{R}$ satisfies the conditions of a statistical complexity measure of being 0 in the thermodynamic limit for the Erd\"os-R\'enyi random graph.
It is known (see \cite{Smith2017a}) that if a graph is regular, that is, every node has the same degree, then the hierarchical complexity is~$0$-- since we have $n$ nodes of degree $k$ all with neighbourhood degree sequences $\{k,k,...,k\}$.

We can also show that for the other end of the entropic spectrum, Erd\"os-R\'enyi random graphs, that the hierarchical complexity tends to 0, as $n$ tends to infinity. To do so we need a formula for the standard deviation of the $i$-th largest sample from this distribution, this is known as the $i$-th order statistic, see \cite{David04} for background on order statistics.
We can use known results on order statistics to derive the following:

\begin{theorem}\label{thm:genER}
    Fix $d\in[0,1]$. The normalised hierarchical complexity of an Erd\"os-R\'enyi graph $\ER(n,d)$ tends to $0$ as $n$ tends to infinity, that is,
    $$\lim_{n\rightarrow\infty}\hat{R}(\ER(n,d))=0.$$

\begin{proof}
For brevity let $\hat{R}_{n,d}:=\hat{R}(\ER(n,d))$.
 Note that if $d=0$ or $d=1$, then $\ER(n,d)$ is regular, thus $\hat{R}_{n,d}=0$ for all $n$, and the result holds. So assume $d\in(0,1)$.
 We begin by giving an approximation of $\hat{R}_{n,d}$. 

First note that the node degrees of an Erd\"os-R\'enyi graph are sampled from the binomial distribution ${B(n-1,d)}$. Whilst this sampling is not strictly independent, the dependence is very weak, and the correlation tends to zero as $n$ tends to infinity \cite{ATB23}, and thus can be disregarded in our approximation and asymptotic analysis.

In \cite[Theorem 9.1]{Baglivo05} an approximate formula for the standard deviation of the $i$-th order statistic of $k$ samples for a continuous distribution with PDF $\phi(x)$ and CDF $\Phi(X)$ is given by:
\begin{align}\label{eq:genapprox}\sigma_i\approx \frac{1}{\phi(\Phi^{-1}(\frac{i}{k+1}))}\sqrt{\frac{i(k-i+1)}{(k+1)^2(k+2)}}.\end{align}
By \cite[Equation (1.1)]{Nag92} this approximation also holds for discrete distributions. 

So to get the $k$'th normalised hierarchical complexity of $\ER(n,d)$, we let $X$ be the binomial distribution ${B(n-1,d)}$, sum \eqref{eq:genapprox} across $i=1,\ldots,k$ and divide by our normalisation, which gives:
\begin{align*}
\hat{R}_{k}(\ER(n,d))&\approx\frac{\sum_{i=1}^k\frac{1}{\phi(\Phi^{-1}(\frac{i}{k+1}))}\sqrt{\frac{i(k-i+1)}{(k+1)^2(k+2)}}}{(1-d)m}\\
&=\frac{2\sum_{i=1}^k\frac{\sqrt{i(k-i+1)}}{\phi(\Phi^{-1}(\frac{i}{k+1}))}}{d(1-d)n(n-1)(k+1)\sqrt{k+2}}.
\end{align*}
The second equality is because the expected number of edges is $m=\frac{n(n-1)d}{2}$. 

We can approximate the expected smallest and largest degree in $\ER(n,d)$, using [Equation 4.5.1]\cite{David04} which gives an approximation for the $i$'th order statistic as $\Phi^{-1}(\frac{i}{n+1})$, for sufficiently large $n$. This gives our lower and upper summands $a$ and $b$, and the global formula is given by averaging $\hat{R}_k$ between $a$ and $b$.

So for large $n$ the global hierarchical complexity can be approximated by
\begin{equation}\label{eq:ER}
    \hat{R}(\ER(n,d))\approx\frac{2\sum_{k=a}^{b}\frac{\sum_{i=1}^k\frac{\sqrt{i(k-i+1)}}{\phi(\Phi^{-1}(\frac{i}{k+1}))}}{(k+1)\sqrt{k+2}}}{(b-a)d(1-d)(n-1)n},
\end{equation}
where $\phi$ and $\Phi$ are the PMF and CDF, respectively, of the binomial distribution $B(n-1,d)$, and $a=\lfloor\Phi^{-1}(\frac{1}{n})\rfloor$ and $b=\lceil\Phi^{-1}(\frac{n-1}{n})\rceil$. Note that this approximation is not particularly close, particularly for small $n$, but it is sufficient to consider the limit of the complexity as $n$ grows.

The binomial distribution can be approximated by the normal distribution, for which the quantile function $\Phi^{-1}$ is known in terms of the inverse error function. Combining this with an approximation of the inverse error function \cite[Equation 13]{Str68} we get:
 \begin{align}\Phi^{-1}(x)&\approx((n-1)d+\sqrt{2(n-1)d(1-d)}\text{ erf}^{-1}(2x-1)\nonumber\\
     &\approx(n-1)d+\sqrt{-\ln(4x(1-x))2(n-1)d(1-d)}\label{invphi}
\end{align}
Combining Equation \eqref{invphi} with the De Moivre-Laplace approximation of the binomial PMF we get:
\begin{align}
    \phi&(\Phi^{-1}(x))\approx\frac{\exp\left(-\frac{(\Phi^{-1}(x)-(n-1)d)^2}{2(n-1)d(1-d)}\right)}{\sqrt{2\pi (n-1)d(1-d)}}\nonumber\\
    &\approx\frac{\exp\left(-\frac{((n-1)d+\sqrt{-\ln(4x(1-x))2(n-1)d(1-d)}-(n-1)d)^2}{2(n-1)d(1-d)}\right)}{\sqrt{2\pi (n-1)d(1-d)}}\nonumber\\
    &=\frac{\exp\left(\ln(4x(1-x))\right)}{\sqrt{2\pi (n-1)d(1-d)}}
    =\frac{4x(1-x)}{\sqrt{2\pi (n-1)d(1-d)}} \label{phiphi}
\end{align}
We can then use Equation \eqref{phiphi} to show that asymptotically $\hat{R}_{n,d}$ is bounded above by zero:
\begin{align}
    \hat{R}_{n,d}&\approx\frac{2}{(b-a)d(1-d)n(n-1)}\sum_{k=a}^{b}\frac{\sum_{i=1}^k\frac{\sqrt{i(k-i+1)}}{\phi(\Phi^{-1}(\frac{i}{k+1}))}}{(k+1)\sqrt{k+2}}\nonumber\\
    &\approx\sum_{k=a}^{b}\frac{2\sum_{i=1}^k\frac{\sqrt{2\pi (n-1)d(1-d)}\sqrt{i(k-i+1)}}{4\frac{i}{k+1}(1-\frac{i}{k+1})}}{(b-a)d(1-d)n(n-1)(k+1)\sqrt{k+2}}\nonumber\\
    &=\sum_{k=a}^{b}\frac{\sqrt{\pi}\sum_{i=1}^k\frac{k+1}{\sqrt{i(k-i+1)}}}{n(b-a)\sqrt{2d(1-d)(n-1)(k+2)}}\nonumber\\
    &=\frac{\sqrt{\pi}\sum_{k=a}^{b}\sum_{i=1}^k\frac{k+1}{\sqrt{i(k-i+1)(k+2)}}}{n(b-a)\sqrt{2d(1-d)(n-1)}}\nonumber\\
    &\le\frac{\sqrt{\pi}\sum_{k=a}^{b}\sum_{i=1}^k 1}{n(b-a)\sqrt{2d(1-d)(n-1)}}\nonumber\\
    &\le\frac{\sqrt{\pi}(b-a)b}{n(b-a)\sqrt{2d(1-d)(n-1)}}\xrightarrow[n\rightarrow\infty]{}0\label{final}
\end{align}

The final step follows since $d$ is fixed and $b$ is always smaller than $n$, thus the denominator grows at least $\sqrt{n-1}$ faster than the numerator.

\end{proof}
\end{theorem}

We conjecture that a similar result holds for random geometric graph (RGG) (see Section~\ref{sec:NetMod} for the definition of RGG's) with a fixed average degree of~$b$ (note in this case, $d\rightarrow0$ as $n\rightarrow \infty$). In particular, when we randomly position $n$ nodes on the unitary Euclidean plane and connect two nodes whenever they are within radius $r=\sqrt{b/(n\pi)}$ of each other. This radius $r$ is selected to ensure we obtain a graph with the required average degree.
In this version of an RGG, the degree distribution is also the binomial distribution (see \cite{Antonioni2012}), so we can apply a similar technique as used in the proof of Theorem~\ref{thm:genER}. However, RGGs have non-trivial degree correlations violating statistical assumptions used in the Erd\"os-R\'enyi case \cite{Antonioni2012}. Nevertheless, we conjecture that a different bound can be obtained that does tend to $0$, we leave this as an open problem.
\begin{conjecture}
Let $RGG(n,r)$ be the random geometric graph in the unitary plane with $n$ nodes and radius $r=\sqrt{\frac{b}{n\pi}}$, for a fixed $b\in\mathbb{N}$. We have
    $$\lim_{n\rightarrow\infty}\hat{R}(RGG(n,r))=0.$$
\end{conjecture}

We've seen that the formula for hierarchical complexity is closely linked to order statistics and the quantile function of probability distributions. Very few closed formulas exist for quantile functions, and they are known to be difficult to analyse. Due to this we are otherwise limited in our analytical treatment of this measure. The remainder of the paper shows the validity of this normalisation in application and we use it to derive novel insights from models and real data. Particularly, we pursue the hypothesis that statistical complexity arises naturally through a combination of hierarchical and geometric components to network connectivity. 

\section{Methods}

\subsection{Network models}\label{sec:NetMod}
Our understanding of this normalisation is aided by its application to different network models and studying the behaviour of our normalisation of HC as we change the size and density of the network. In the following, a graph refers to a mathematical object of a set of nodes with adjoining edges. A network refers to a graph representation of the relationships or connections between components of a real-world complex system.

Firstly, we use \textbf{Erd\"os-R\'enyi (E-R) Random Graphs}. {E-R} random graphs are generated using random uniform edge probabilities in $[0,1]$ \cite{Erdos1959}.
They give an indication of the behaviour of an `average' graph of a given size and density. That being said, they do not give any indication of the behaviour of an `average' network as it lacks many of the basic characteristics common to networks such as a relatively high clustering coefficient and degree heteregoneity.

\textbf{Random Geometric Graphs (RGG)} are generated from randomly sampled co-ordinates in the unit cube (i.e. 3D) \cite{Dall2002}. These samples then represent nodes and the inverse distances between node pairs are the weights of the edges between them. For a desired network density, we select the $m$ largest weights as our graph edges. RGGs have properties of high clustering desirable for networks, however they also lack the characteristic degree heterogeneity of networks.

Surface-Depth (S-D) models provide geometric graphs with heterogeneous degree distributions which show distinct similarities to many real-world networks \cite{Smith2021}. We shall refer to these models throughout as \textbf{Random Hierarchical Geometric Graphs (RHGG)} since they combine a geometric component to connections with a hierarchical component, both of which we also want to study in isolation. These are generated using two parameters, the $\sigma$ of a log-normal distribution and the number of dimensions, $q$, of a random geometric graph. The weights of the edges are then defined as 
\begin{equation}
w_{ij} = d_{ij}(s_{i}+s_{j}),
\end{equation}
where $d_{ij}$ is the inverse distance between nodes $i$ and $j$ in a random geometric graph with $q$ dimensions and each $s_{i}$ is a random sample from a log-normal distribution $LN(\mu,\sigma)$. Again, for a desired network density, we select the $m$ largest weights as our graph edges. In this study, we perform a basic exploration with $q=3$, $\mu = 0$, and $\sigma$ fixed at 0.2 as this produces graphs with a suitable heterogeneity. We then also explore the effect of varying $\sigma$, and so heterogeneity of geometric graphs, on hierarchical complexity. 

Configuration models of these RHGGs allow us to probe the extent to which hierarchical complexity of the RHGGs can be attributed merely to the hierarchical structure of the network. We refer to these as \textbf{Random Heterogeneous Graphs (RHGs)} to emphasise the relationship with our other models. Briefly, configuration models work by fixing the degree distribution of a network but otherwise randomising the connections. Each node is provided with a number of stubs equal to its degree. These stubs are then randomly paired between nodes to establish edges \cite{Sneppen2002}.

\subsection{Link growth mechanisms}
We used ten different link growth mechanisms to observe the effect on NHC of increasing density in different ways on real-world networks.

For all node pairs without edges, $(i,j)\notin \mathcal{E}$ we considered:

\subsubsection*{\textbf{Random growth}} edge probabilities are uniform.

\subsubsection*{\textbf{Popularity growth}} similar in fashion to preferential attachment for addition of new nodes to a network, the probability of a new edge is proportional to the sum of the degrees of the nodes:
\begin{equation}
    p_{ij}\sim k_{i} + k_{j}.
\end{equation}

\subsubsection*{\textbf{Similarity growth}} for $g_{i}, g_{j}$ the neighbourhoods of nodes $i$ and $j$, the probability of edge $(i,j)$ occurring is proportional to the Jaccard index of their neighbourhoods:
\begin{equation}
    p_{ij}\sim J(g_{i},g_{j}) = \frac{|g_{i}\cap g_{j}|}{|g_{i}\cup g_{j}|}.
\end{equation}

\subsubsection*{\textbf{Popularity $\times$ similarity growth}} the probability of a new edge is proportional to the intersection of their neighbourhoods, essentially removing the size normalisation of the Jaccard index:
\begin{equation}
    p_{ij}\sim |g_{i}\cap g_{j}|
\end{equation}
so that the probability of connection is dependent on the size of the neighbourhoods and the overlap of the neighbourhoods. Note, popularity and similarity growth is what is commonly known as the common neighbours algorithm, as it just counts the number of shared neighbours two nodes have to predict whether they will become connected in the future \cite{Newman2001,Cannistraci2013}. While simple at face value, we can clearly see from the above that this works well to account for both popularity and similarity components of a network.

For each of the three latter approaches (popularity, similarity, and popularity $\times$ similarity) we took three different approaches to deciding on links: random probabilistic selection, random exponentiated probabilistic selection, and deterministic rank-based selection.

\subsubsection*{Probabilistic} To get the probabilities, $p_{ij}$, we divide each individual measurement (e.g., $k_{i}+k_{j}$ for hierarchical attachement) by the sum over all available measurements for example, 
\begin{equation}
    p_{ij} = (k_{i}+k_{j})/\sum_{(i,j)\notin\mathcal{E}}(k_{i}+k_{j}). 
\end{equation}
Edges are then randomly selected based on these probability spaces $\{p_{ij}\}_{(i,j)\notin\mathcal{E}}$.

\subsubsection*{Exponentiated probabilistic} This approach is taken to better differentiate between the strong and weak potential links in the probability space and so make it more likely for stronger potential links to be selected. Here, we simply take the exponentials of the probabilities before normalisation: $p_{ij} = \exp(k_{i}+k_{j})/\sum_{(i,j)\notin\mathcal{E}}\exp(k_{i}+k_{j})$.

\subsubsection*{Deterministic} Here, we simply take the top $x$ probabilities as new links. This is typically how link prediction would be done.

These three approaches span from the more randomised, to the more rigid, with exponentiated probabilistic growth taking the middle ground.

\subsection{Data}
We obtained data for twenty large networks from two databases-- the SNAP database \cite{snapnets} and the Network Repository 
\cite{nr}. While hierarchical complexity has been applied to several different types of networks (including social networks, protein networks and infrastructure networks) with mixed results compared to configuration models \cite{Smith2019b}, it has yet to be applied to larger sized networks. Further, we have so far been unable to directly compare hierarchical complexity of networks of different sizes due to the lack of a normalisation.

Network were chosen to cover a wide range of sizes (1912-36692) and types (protein interaction networks, social networks, infrastructure networks, collaboration networks), and also to include groups of certain types of networks to explore relationships within and between network types. The number of nodes, edges and the network density for each network are shown in Table \ref{tab:NetStats}.

Protein-protein interaction networks generated from co-expression correlations were taken from the Network Repository, which in turn derived these graphs from data from wormnet \cite{cho2014wormnet}. These were obtained for Homo Sapiens (HS), Derio Rario (DR)-- zebrafish, Drosophilia Melanogaster (DM)-- fruit fly, and caenorabditis elegans (CE)-- a nematode. All of these are exceptionally well studied, model species for which the data is most extensive and reliable.

Also from the Network repository we took two infrastructure networks, one being the widely studied network of the Western States power grid of the US (power grid) \cite{watts1998} and the other being a network of international flights between airports where nodes are airports and edges are established where there are flights between those airports (open flights) \cite{opsahl2011}. All other networks were obtained from the SNAP repository. 

We studied five collaboration networks within Physics disciplines, constructed from arXiv data. For these, edges are established between co-authors of papers. Topics are self-selected by authors during arXiv manuscript uploads. These topics are astrophysics (collab AstroPh), condensed matter physics (collab CondMat), general relativity (collab GrQc), high energy physics (collab HepPh), and high energy physics theory (collab HepTh).

We studied six social networks constructed from the twitch platform for six different languages-- English (twitch ENGBE), French (twitch FR), German (twitch DE), Portuguese (twitch PTBR), Russian (twitch RU), and Spanish (twitch ES) \cite{Rozemberczki2019}. Nodes are the users of twitch and edges are friendships between them. Facebook page-page is another social network where nodes represent official Facebook pages while edges are mutual likes between sites, collected through the Facebook Graph API in November 2017 and restricted to four categories of pages- politicians, governmental organisations, television shows and companies \cite{Rozemberczki2019}. The LastFM social network is the network where nodes are users from Asian countries and edges are mutual follower relationships between them \cite{Rozemberczki2020}. We also analysed a large email network between Enron employees (email Enron), originally made public by the Federal Energy Regulatory Commission during its investigation. Here, nodes are email addresses and edges are established wherever there are emails sent between addresses.

\begin{table}
    \centering
    \caption{Statistics for twenty real world networks}
    \begin{tabular}{|r|c|c|c|}
    \hline
          \textbf{Network}     & $n$ & $m$ & $d$ \\
         \hline
                  email Enron & 36692 & 367662 & 0.0005 \\
         \hline
                  Facebook & 22470 & 171002 & 0.0007 \\             
         \hline
                  collab CondMat & 22167 & 186936 & 0.0008 \\       
         \hline
                  collab AstroPh & 16000  & 396160 & 0.0031 \\      
         \hline
                  protein CE & 15229  & 245952 & 0.0021 \\          
         \hline
                  collab HepPh & 12008  & 237010 & 0.0033 \\        
         \hline
                  collab HepTh & 9877  & 51971 & 0.0011 \\          
         \hline
                  twitch DE & 9498  & 153138 & 0.0034 \\            
         \hline
                  lastFM Asia & 7624  & 27806 & 0.0010 \\           
         \hline
                  twitch ENGBE & 7126  & 35324 & 0.0014 \\          
         \hline
                  twitch FR & 6549  & 112666 & 0.0053 \\            
         \hline
                  collab GrQc & 5242  & 28980 & 0.0021 \\           
         \hline
                  power grid & 4941  & 6594 & 0.0005 \\             
         \hline
                  twitch ES & 4648  & 59382 & 0.0055 \\             
         \hline
                  protein HS & 4413  & 108818 & 0.0112 \\           
         \hline
                  twitch RU & 4385  & 37304 & 0.0039 \\             
         \hline
                  protein DM & 4040 & 76717  & 0.0094 \\            
         \hline
                  protein DR & 3289 & 84940 & 0.0157 \\             
         \hline
                  open flights & 2939 & 30501 & 0.0071 \\           
         \hline
                  twitch PTBR & 1912 & 31299 & 0.0171 \\            
         \hline
    \end{tabular}
    \label{tab:NetStats}
\end{table}

\subsubsection*{Allen brain model for use in large network experiment}
For studying the effect of normalisation in very large graphs, we used the mouse V1 model from the Allen Institute for Brain Science \cite{Billeh20}. This model contains approximately 230,000 neurons, and can be considered as a network where each neuron is a node and there is an edge between two nodes if they are connected by a synapse. We can sample geometric cylinders from this model using the associated code provided at \cite{v1_model}, this allows us to construct networks that should be very similar structurally, but vary in size from anything up to 230,000 nodes. Due to computational limitations we only compute complexity values on up to 95,000 nodes.

\subsubsection*{Data for replication in link growth mechanism experiment}
For studying the link growth mechanisms, we used a replication dataset from the ICON corpus consisting of 139 networks mostly describing biological, social and technological phenomena \cite{Ghasemian2018, Smith2019b}. These ranged from $n = 50$ to $n = 3155$ with a mean of  341 $\pm$ 462. Densities ranged from $d= 0.0011$ to $d= 0.3884$ with a mean of 0.0578 $\pm$ 0.0717.

\section{Results}\label{sec:results}

\subsection{Normalisation results on random models}
 
\begin{figure*}[!t]
\centering
\includegraphics[trim = 0 0 0 0, scale = 0.34]{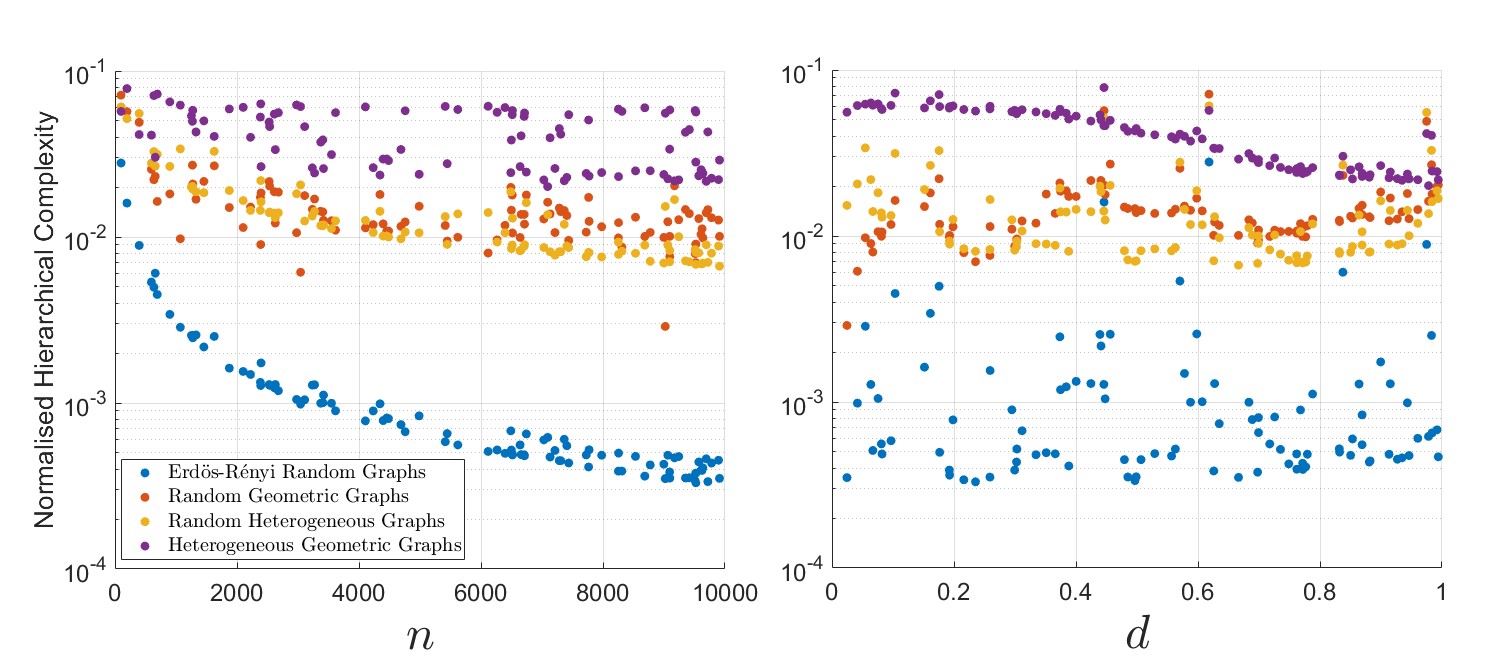}
\caption{Results show measurements for random realisations ($n\sim U[50,10000]$, $d\sim U[0,1]$) of different random graphs as denoted in the legend, against size $n$ and density $d$.}
\label{fig_samplingError}
\end{figure*}

Fig. \ref{fig_samplingError} shows the results of the normalisation applied to our chosen random graph models. For each random model (Erd\"os-R\'enyi, RGG, RHG, and RHGG) we generated 100 realisations with $n\sim U[50,10000]$ and $d\sim U[0,1]$ and plot results against both $n$ and $d$.

Notably, Erd\"os-R\'enyi random graphs have the lowest complexity of the models studied. Above this, the complexity of RGGs and RHGs is similar across all $n$ and $d$. The greatest complexity is clearly observed in the RHGGs, particularly at lower densities, indicating that greater complexity arises naturally through the combination of hierarchical and geometric structure.

While Erd\"os-R\'enyi random graphs, RGGs and RHGs have similar levels of complexity across density (highlighting the normalisation features of our measure across density), there is a different behaviour noted in RHGGs. Higher complexity at low densities compared to high densities in this instance can be considered a structural feature present in the graphs. We can understand very high densities as regarding the connectivity of the least important connections (specifically, the complement of the graphs) which in this instance are those between nodes with low degrees which are geometrically distant. It is reasonable to expect complexity here to be as low as for RHGs and RGGs.

As per the theoretical results for E-R random graphs, the experiments across the different models shows an inverse relationship for $\hat{R}$ with $n$ but little to no dependency on $d$. The reason for hierarchical complexity decreasing with increasing $n$ may be a true relationship of the complexity of these models as $n$ increases, rather than a normalisation issue. Indeed, we can expect that larger sample sizes of neighbourhood degree sequences given by larger random graphs would result in more homogeneous ordered sequences as they better approximate the global degree distributions. 

Interestingly, for the RHGGs-- which more accurately model real-world network structure-- there is very little if any decrease with increasing $n$ for $n>1000$. This indicates that comparisons of $\hat{R}$ in large networks are reliable, but caution must still be taken when making comparisons in smaller networks.

In the Supplementary Information section SI1,  we explore the suitability of two relevant previously proposed network measures for assessing statistical complexity of networks using similar analyses as above. We find that neither meet the requirements expected of being 0 for regular and ER random graphs. We also find that they are not useful for distinguishing between different types of models which should be distinguishable in terms of statistical complexity.

\subsection{Normalisation results on increasing number of nodes}
To extend our observations of the normalisation with respect to network size we studied the behaviour of the normalisation in very large graphs. In this case we use the mouse V1 model from the Allen Institute for Brain Science, see \cite{Billeh20}, and Erd\"os-R\'enyi random graphs, for which we know the normalised complexity value tends to $0$ by Theorem~\ref{thm:genER}. 

We would expect networks of the same structure to have the same, or at least similar, NHC. However, what do we mean by the ``same structure"? If two networks have a different number of nodes they inherently have a different structure. In fact, the size of the network is related to the complexity, because as the size of random graphs increase their uniformity increases, this is due to the inverse relationship between variance and the sample mean. As such, we would expect that the NHC will decrease slightly as the number of nodes increases, but two sufficiently large networks of similar structure, but different size, would have similar complexity values.

\begin{figure}[!t]
\centering
\includegraphics[width=5in]{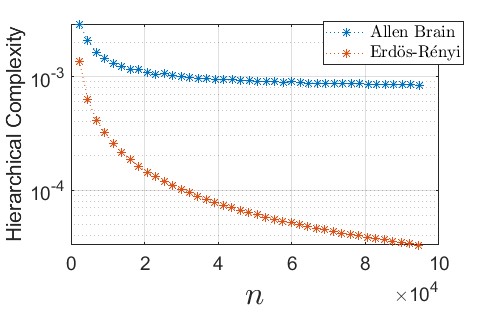}
\caption{The normalised hierarchical complexity of cylinders of increasing sizes of the Allen Brain V1 mouse model vs ER graphs of the same size and density.}
\label{fig_allen}
\end{figure}

The effect of this is well demonstrated in Fig~\ref{fig_allen}. We see that the Erd\"os-R\'enyi random graph tends to 0 with increasing $n$, as expected. At the same time, for the Allen Brain for small $n$ the complexity is higher, but for larger $n$, roughly $n\ge2000$, the complexity is very close between samples, appearing to tend to a non-zero limit in $n$. This demonstrates a behaviour expected of statistical complexity in dynamical systems, that randomness (aswell as regularity) vanishes to 0 complexity in the limit of $n$, while non-zero complexity is maintained in diverse structure \cite{Feldman1998}.

\subsection{Effect of degree heterogeneity on hierarchical complexity}
\begin{figure}[!t]
\centering
\includegraphics[width = 5in]{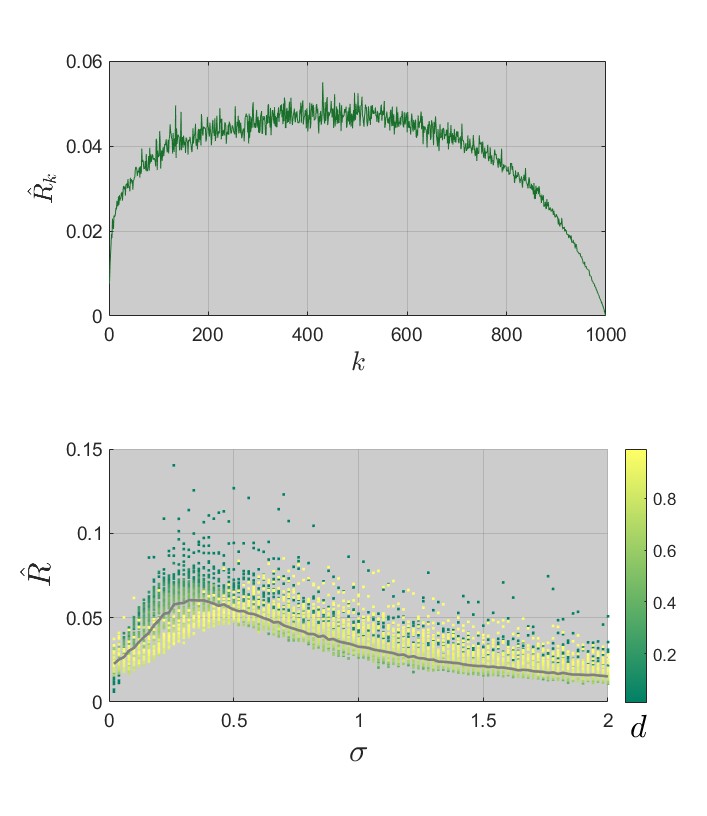}
\caption{Top, average normalised hierarchical complexity per degree of heterogeneous geometric graphs with with $n=1000$, $d\sim U[0,1]$, and $\sigma_{h}= 0.01,0.02,\dots,2$. One hundred realisations are created for each $\sigma$. Bottom, global normalised hierarchical complexity plotted against heterogeneity of these networks.}
\label{fig_heterog_sigma}
\end{figure}

Here we study the change in complexity as we increase heterogeneity among the RHGGs. We generated RHGGs of size $n=1000$ and $d\sim U[0,1]$. The heterogeneity was determined with 100 realisations of the model for each $\sigma_{h} = 0.01,0.02,\dots,2$. Results are shown in Fig~\ref{fig_heterog_sigma}.

The results show that NHC has a consistent range of values across most degrees for heterogeneous geometric graphs, tapering off quickly at either end towards 1 and $n$. It is beneficial that the measure does not have a positive or negative relationship with degree so that the measure does not overly emphasise any particular range of degrees. Further, the fact the measure quickly tapers off towards very small degrees protects the measure from being influenced by the high levels of uncertainty of sampling at these small sample sizes.

NHC is generally strongest in models with $\sigma_{h}$ between 0.2 and 0.4 and decreases towards low and high heterogeneity. When $\sigma_{h}$ is high the network becomes dominated by the hierarchical relationships, which should make the network more ordered. On the other hand, with low $\sigma_{h}$ the model gets closer to a random geometric graph which has low values of NHC as shown in Fig~\ref{fig_samplingError}. This is consistent with our expectations of NHC being a statistical measure of complexity in networks.

In section SI2 of the Supplementary Information, we apply NHC to the non-uniform Popularity Similarity Optimisation model \cite{Muscoloni2018} to see how it fairs on another model which directly utilises degree heterogeneity and latent geometry. We find similar patterns as for RHGGs, that the model achieves highest NHC in a middle ground of degree heterogeneity. We also explore the effect of varying the clustering in the network and the number of communities through additional parameters of the model. We find that decreasing the clustering by randomising more connections decreases NHC, and that introduction strong community blocks in the geometry also decreases NHC. These are broadly consistent with our expectations of a statistical complexity metric.

\subsection{Hierarchical complexity in large real-world networks}
We compared unnormalised and NHC in the twenty large networks detailed in Table \ref{tab:NetStats}. The results are shown in Table \ref{tab:HCrankings}. The values of the non-normalised HC range from 4 in the power grid network to 87858 in the German twitch network, a difference by a factor of nearly 22000. This range highlights the lack of utility of an unnormalised and unbounded measure for comparisons of different networks. On the other hand, the values of NHC range from 0.0022 in the power grid network up to 0.0830 in the Portugese twitch network, a difference by a factor of just under 38.

We can see that the normalised measure allows us to compare between these networks more clearly. For example, consider the twitch DE, twitch FR and twitch RU networks, which have the largest non-normalised HC values. These three networks are all constructed in the same way (friendships in twitch) and have similar densities (0.0034,0.0053, and 0.0039, respectively), and yet the twitch DE non-normalised HC value is 6 times bigger than the twitch RU non-normalised HC, suggesting that the twitch DE network is significantly more complex than the twitch RU network. However, when normalised these three networks all have very similar NHC values, suggesting they have very similar levels of complexity, as one would expect.

Next consider the other three twitch networks: ES, PTBR and ENGBE. We can see that twitch ES and twitch ENGBE have very similar unnormalised HC values, but when normalised the twitch PTBR is twice as complex as twitch ES. This is not surprising considering they have markedly different densities, and we will see in the next section that even after normalisation the network density is correlated to the NHC value in real world networks. We also see that the twitch ENGBE has much lower complexity than the other twitch networks, which is likely explained by the lower density of the network.

So we can see that our normalisation reduces the scale of the difference between complexities of networks, and allow us to better compare networks of different sizes. However, we also see that the density of the network still correlates with the NHC value.

\begin{table}
    \centering
    \caption{Rankings of hierarchical complexity and normalised hierarchical complexity for twenty real world networks}
    \begin{tabular}{|r|c||c|c|}
    \hline
          $R$     & \textbf{Network} & $\hat{R}$ & \textbf{Network}\\
         \hline
                  87858     & twitch DE         & 0.0830 & twitch PTBR\\        
         \hline
                  35996     & twitch FR         & 0.0684 & twitch RU\\          
         \hline
                  14328     & twitch RU         & 0.0612 & twitch DE\\              
         \hline
                  11099     & email Enron       & 0.0586 & twitch FR\\        
         \hline
                  7625      & twitch ES         & 0.0526 & open flights\\        
         \hline
                  7167      & twitch PTBR       & 0.0454 & twitch ES\\         
         \hline
                  4216      & twitch ENGBE      & 0.0360 & protein DR\\                  
         \hline
                  3275      & facebook          & 0.0334 & protein HS\\       
         \hline
                  2207      & collab HepPh      & 0.0297 & email Enron\\
         \hline
                  1283      & protein HS        & 0.0279 & protein DM\\ 
         \hline
                  1015      & protein DR        & 0.0272 & twitch ENGBE\\
         \hline
                  873       & collab AstroPh    & 0.0258 & collab HepPh\\ 
         \hline
                  755       & protein CE        & 0.0182 & facebook\\                 
         \hline
                  752       & protein DM        & 0.0154 & collab GrQc \\ 
         \hline
                  582       & open flights      & 0.0147 & lastFM Asia\\                
         \hline
                  293       & LastFM Asia       & 0.0124 & protein CE\\        
         \hline
                  261       & collab CondMat    & 0.0105 & collab AstroPh\\               
         \hline
                  68        & collab GrQc       & 0.0056 & collab HepTh\\
         \hline
                 39         & collab HepTh      & 0.0054 & collab CondMat\\
        \hline
                  4         & power grid        & 0.0022 & power grid\\
         \hline
    \end{tabular}
    \label{tab:HCrankings}
\end{table}

\subsection{Growth of hierarchical complexity in real world networks}
\begin{figure*}[!t]
\centering
\includegraphics[width=7in]{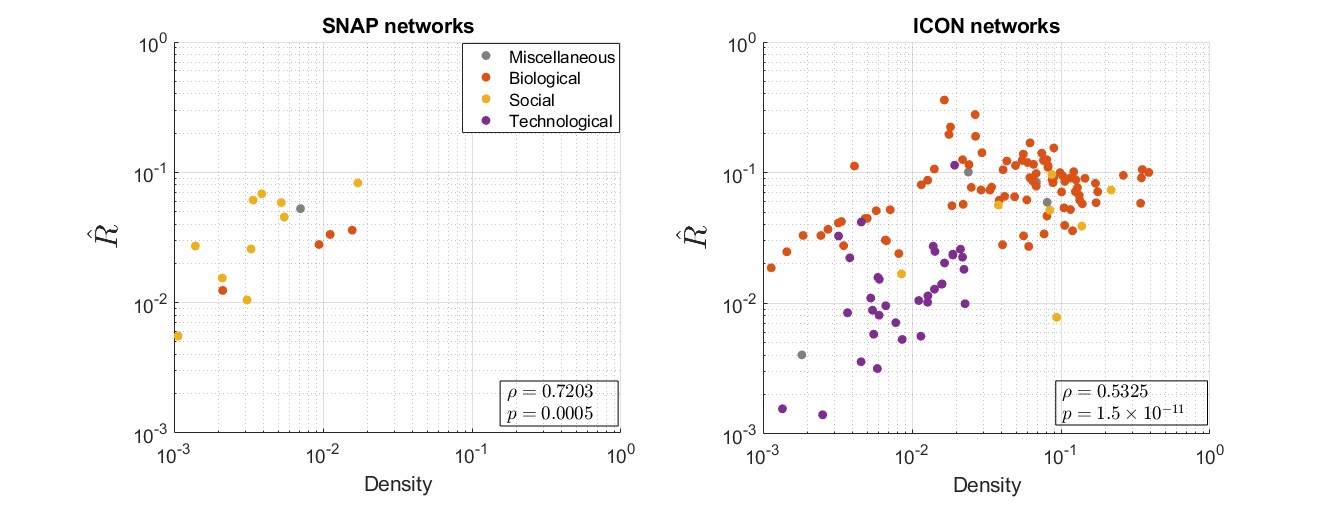}
\caption{Scatterplots visualising the positive association between density and hierarchical complexity in real-world networks. Spearman's correlation coefficient and associated $p$-value shown inset. Bottom row shows average results of the values of NHC as we increase density of networks according to the link growth mechanisms as described in the legend.}
\label{fig:corr}
\end{figure*}

We found that hierarchical complexity was positively correlated with network density in the twenty networks from Table~\ref{tab:NetStats} ($\rho = 0.7203$, $p = 0.0005$), Fig. \ref{fig:corr}, left. We confirmed this observation with a network dataset of 139 smaller networks ($n\in[50,3155]$, $\rho = 0.5325$, $p= 1.5\times10^{-11}$) \cite{Ghasemian2018,Smith2019b}, Fig. \ref{fig:corr}, right. To discount the potential confounding effect of network size, $n$, on these correlations, we implemented linear regression on network density with network size as a predictor and found that the correlations of the residuals of the regression with NHC were still significant in both cases-- $\rho = 0.6526$, $p=0.0023$ for the 20 large networks and $\rho = 0.2294$, $p=0.0066$ for the 139 small-to-medium sized networks.

At the same time, we have shown that NHC shows strong normalisation with respect to density for many types of graph. The relationship between density and NHC in real world networks is therefore unlikely due to a lack of normalisation, but is a true relationship requiring a mechanistic explanation. To try to explain this relationship we applied ten link growth algorithms, as described in section III.B, to real-world networks to artificially increase their density and see if any would consistently lead to the targeted increase in NHC.

\begin{figure*}[!t]
\centering
\includegraphics[trim = 50 0 0 0,width=7in]{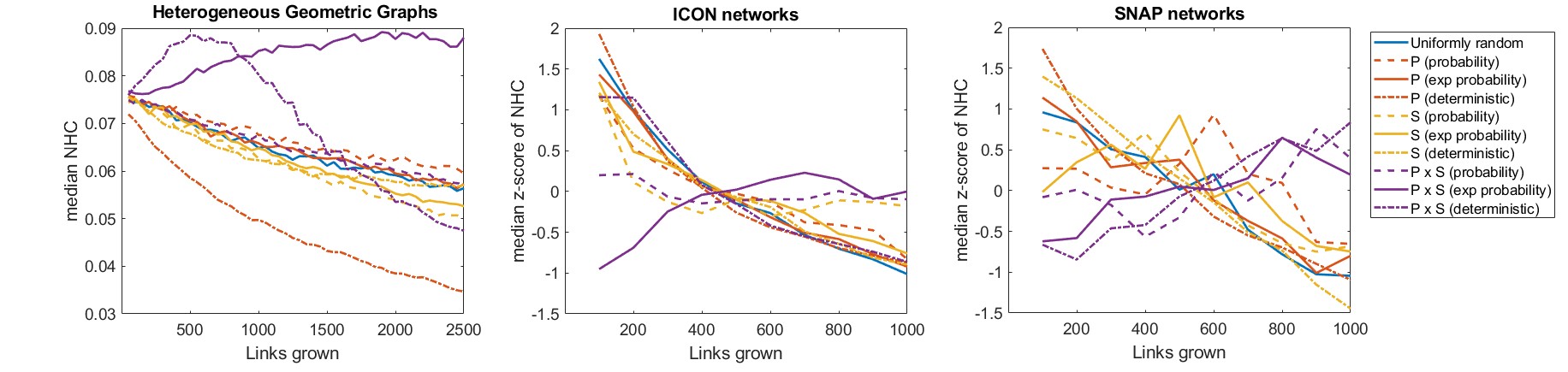}
\caption{Average results of the values of NHC as we increase density of networks according to the link growth mechanisms as described in the legend. P stands for Popularity, S for Similarity and P $\times$ S for Popularity $\times$ Similarity. For the Heterogeneous geometric graphs we take the median over 100 iterations. For the ICON and SNAP networks we take the median of the z-scores of trajectories to avoid bias of networks with high NHC.}
\label{fig:grow}
\end{figure*}

Firstly, we implemented link growth on 100 HRGGs of size $n=250$, density $d=0.05$ and density $\sigma = 0.2$. Each link growth algorithm in Section III.B was applied for 50 iterations with 50 new links established each iteration. For the real-world networks, we applied the link growth algorithms to all of the 139 ICON networks and 17 of the 20 SNAP networks (omitting the 3 largest due to computational limitations). For these we used 10 iterations with 100 new links established each iteration. Results are shown in Fig. \ref{fig:grow}.

There are two clear observations to be made from these results. Firstly, the only kind of link growth which provided increased NHC were those based on the common neighbours (here referred to as Popularity $\times$ Similarity) algorithm. This shows a clear favoured strategy for link growth in complex networks.

All of the random, Popularity only and Similarity only algorithms worked to decrease NHC and in similar amounts as the baseline uniformly random algorithm. There was no particular observable consistent difference between the Popularity and Similarity only mechanisms. However, the deterministic mechanisms typically did worse most of the time. Infact, in the HRGGs the Popularity only deterministic mechanism did much worse than the uniformly random approach. Anytime the algorithm showed a greater decrease than the uniformly random mechanism we can assume this is caused by a more ordered structure being enforced on the network, since the only NHC smaller than ER random graphs are regular graphs and highly organised graphs where all nodes of the same degree have the same or similar neighbourhood degree sequences \cite{Smith2019b}.

Secondly, the only algorithm which consistently increased the NHC across the three datasets was the exponentiated probabilistic Popularity $\times$ Similarity algorithm. This algorithm takes the middle ground between the more random standard probabilistic algorithm and the deterministic link ranking algorithm. The probabilistic mechanism increased for the SNAP networks over 10 iterations, but decreased slightly in the ICON networks and moreso in the RHGGS with a similar trajectory as for uniformly random growth. The deterministic mechanism intially increased alot for the RHGG model before going into a steep decline, while it increased in the SNAP networks but decreased in line with the uniformly random mechanism in the ICON networks.

\section{Discussion}
Our modelling demonstrated greater statistical complexity arising through the combination of hierarchical and geometric components. While the space of random geometric graphs is regular, the nodes being placed randomly in that space opens up pockets of higher and lower connectivity, as quantified through degree correlations \cite{Antonioni2012}. This may be understood as a classic instance where randomness and regularity interact to generate some degree of complexity. Similarly, heterogeneous random graphs contain an ordered structure in terms of the hierarchy of node degrees, combined with randomness of connections established through the configuration model procedure. By giving a randomly allocated log-normal node fitness to the nodes randomly placed in Euclidean space we see an amplification in terms of complexity. This is particularly interesting since heterogeneous geometric networks closely model many aspects of real-world networks \cite{Smith2021}. Future work will explore whether access to heterogeneous connectivity patterns in networks facilitates the formation of more heterogeneous functionality and therefore assess the advantages conferred by the combination of hierarchy and geometry which real-world networks appear to incorporate almost universally, as found in e.g. \cite{Smith2019a} (although there are a subset of networks which buck this trend \cite{Cannistraci2013}).

In the results of NHC among the 20 large real-world networks some patterns appear. The general trends among the three groups of networks we have (6 twitch networks, 4 protein networks, 5 physics collaboration networks) indicate that the twitch networks tend to have highest complexity, while protein networks have fairly high complexity and physics collaboration networks have low complexity. The power grid having the lowest complexity can be expected as it is a network with high geometrical constraints and we might expect some specific universal design principles in its construction. On the other hand, online social networks of twitch are largely free from geometrical constraints (although will still have a latent similarity space, but this can be of arbitrarily large dimension) and may reflect the diversity of social relationships. However, sample sizes would need to be increased to provide stronger evidence for any such generalisations.

The results from the link growth mechanism experiments showed that explanations for the positive relationship between density and NHC were not given by link growth mechanisms for popularity or similarity of nodes separately, but again the combination of the two. It indicates that growth of statistical complexity requires a trade-off of randomness and determinism. Too much randomness and no structure is developed. Too little randomness and the structure becomes too rigid. It also highlights how real-world networks may naturally grow to develop higher statistical complexity simply through nodes more likely to (but not with certainty)) form links with nodes with a lot of common neighbours. It also highlights some amount of futility in trying to ever perfectly predict links-- there is randomness in connectivity, and in fact networks may well benefit from that randomness in breaking rigidity of patterns and becoming more diverse.

The evidence that random geometric graphs have non-zero NHC in the thermodynamic limit gives an interesting insight into the geometrical nature of NHC. It tentatively points towards a definitive notion of statistical complexity of networks. Essentially, if we take seriously the established notion of  networks as embedded in a latent geometrical space \cite{Krioukov2010, SmithA2019}, then we can start to conceptualise measures of statistical complexity of networks such as NHC as attempting to measure the irregularity of the distribution of points over that space.

It is worth noting that in all of the NHC values we computed, we rarely find anything above 0.1. This means, we expect 0.1 is a very high value of NHC for a network. Here, we recall that a normalisation does not require values to be in any particular range, just that the values are comparable for the same phenomena, for example model with all other parameters fixed, with changes to the targeted normalisation parameter. Further, while we provide an upper bound of 2 to demonstrate there is no possible case of the measure exploding to infinity, we do not believe this is a tight upper bound. From our experience, it is likely a tight upper bound is even below 1.

In the future, we intend to thoroughly explore the application of NHC for analysis of brain networks, across scales, across different types of networks, and eventually, across species. We will also explore applications to protein-protein interaction networks utilising the vast datasets available through the STRING database in order to begin to answer questions regarding the relationship between NHC of protein-protein interactions and evolutionary parameters. Applications to social networks hold obvious appeal given the consistently high complexity we noticed in the twitch social networks. There is also serious scope for extension and improvement of NHC. For example, we could consider generalising the measure to considering neighbourhoods of neighbourhoods, up to an arbitrary depth. Generally speaking, our understanding of statistical complexity of networks in different fields is limited by a lack of parsimonious measurements for its quantification. The tools offered here can therefore help researchers to begin to answer fundamental questions regarding the existence and extent of statistical complexity of real-world networks, especially in light of the larger and higher quality datasets becoming evermore available.

\section{Conclusion}
We proposed and demonstrated the utility of a normalisation for hierarchical complexity-- NHC. We proved that this measure is bounded above by $2$ and tends to zero for Erd\"os-R\'enyi random graphs with increasing size. This is analogous to a defining characteristic required of a statistical complexity measure in dynamical systems. We then demonstrated that, while random graph models containing degree heterogeneity and geometry individually had lower complexity, the combined components of degree heterogeneity and geometry is enough to create NHC of a similar level to real-world networks. However, we then found that real-world networks displayed an association between NHC and density that could not be explained solely by our models. 
Instead, we could manage to explain this consistently with a common neighbours link growth algorithm with exponentiated probabilities. Particularly, this was more consistent than a deterministic weight ranking algorithm and the more random non-exponentiated probability algorithm. All other algorithms tried failed to increase complexity. We therefore posit that real-world networks have a preference for growth which increases complexity of the interacting system. We provide a parsimonious measure for statistical complexity of networks which is ready to be applied to answering questions and gathering new insights into the degrees of complexity in various fields such as neuroscience, protein biology and social networks.

\section*{Acknowledgment}
The authors would like to thank Dr April S. Kleppe for her thoughtful feedback on our work.

\ifCLASSOPTIONcaptionsoff
  \newpage
\fi

\bibliography{References}

\end{document}